\documentclass[article,nobalancelastpage]{revtex4}
\usepackage{amssymb}
\usepackage{dcolumn}
\usepackage{graphicx}

\input{tcilatex}

\begin{document}

\title{INVARIANTS\ OF 3D TRANSFORMATION FOR POINT ROTATION COORDINATE FRAMES 
}
\author{Boris V. Gisin}
\affiliation{IPO, Ha-Tannaim St. 9, Tel-Aviv 69209, Israel. E-mail: gisin@eng.tau.ac.il}
\date{\today }

\begin{abstract}
\noindent\ \ \ \ \ \ \ \ \ \ \ \ \ \ \ \ \ \ \ \ \ \ \ \ \ \ \ \ \ \ \ \ \ \
\ \ \ \ \ \ \ \ \ \ \ \ \ \ \ \ \ \ \ \ \ \ \ \ \ \ \ \ \ \ \ \ \ \noindent 
\begin{minipage}[t]{7.2cm}
In Nature there is no so small thing which, by intend look turned to it, 
would not grow up to infinity 
\end{minipage}\ \ 

\ \ \ \ \ \ \ \ \ \ \ \ \ \ \ \ \ \ \ \ \ \ \ \ \ \ \ \ \ \ \ \ \ \ \ \ \ 

\noindent\ \ \ \ \ \ \ \ \ \ \ \ \ \ \ \ \ \ \ \ \ \ \ \ \ \ \ \ \ \ \ \ \ \
\ \ \ \ \ \ \ \ \ \ \ \ \ \ \ \ \ \ \ \ \ \ \ \ \ \ \ \ \ \ \ \ \ \ \ \ \ \
\ \begin{minipage}[t]{5cm}
{\it L. N. Tolstoi }
\end{minipage}\ \ \ \ \ \ \ \ \ \ \ \ \ \ \ \ \ \ \ \ \ \ \ \ \ \ \ \ \ \ \
\ \ \ \ \ \ \ \ \ \ \ \ \ \ \ \ \ \ \ \ \ \ \ \ \ \ \smallskip\ \ \ \ \ \ \
\ \ \ \ \ \ \ \ \ \ \ \ \ \ \ \ \ \ \ \ \ \ \ \ \ \ \ \ \ \ \ \ \ \ \ \ \ \
\ \ \ \ \ \ \ \ \ \ \ \ \ \ \ \ \ \ \ \ \ \ \ \ \ \ \ \ \ \ \ \ \ \ \ \ \ 
\hspace{20cm}
\end{abstract}

\pacs{42.50.Xa, 03.65.Ta, 06.20.Jr, 06.30.Gv }
\maketitle

\section{Introduction}

Recently the general linear transformation for point rotation coordinate
frames was considered \cite{job}. A distinguishing feature of the frame, in
contrast to the Cartesian one, is \ the existence of the rotation axis at
every point. The frame coordinates are an angle and time, the frequency of
rotation is a parameter. The concept of the frame originated from the
optical indicatrix (index ellipsoid) \cite{Kam}. Rotation of the optical
indicatrix arises in three-fold electrooptical crystals under the action of
the rotating electric field applied perpendicular to the optical axis \cite%
{pat}. Such a rotation is possible as in the Pockels as Kerr crystals and
also in the isotropic Kerr medium. The rotation is used in single-sideband
modulators \cite{sm}, \cite{jpc}. The single-sideband modulation has very
interesting features from the theoretical viewpoint. In applications it may
be used for the frequency modulation and frequency shifting. In contrast to
usual modulation such a shifting is "100\% transformation" of the initial
into output frequency. However at present the modulation practically is not
in use. It is connected with the high controlling voltage of bulk modulators 
\cite{jpc}; creating waveguide single-sideband modulators calls for
considerable technological efforts \cite{wssm}.

Two point are essential by the consideration of a plane circularly polarized
wave propagating through a medium with the rotating optical indicatrix. That
is the necessity to use the non-Cartesian point rotation frame and the
necessity to know what is the frequency superposition law by the transition
from one rotating frame to another. The usual description in the Cartesian
frame tacitly assumes that this law is Galilean or linear one. In this law
the frequency of any field may be infinitely large.

In the general case considered in \cite{job} the reverse frequency, i.e.,
the frequency of the second frame relative to the first one, is a function
of the direct frequency. However both the frequencies are assumed to be
symmetric, i.e., the direct frequency is the same function of the reverse
frequency. Using symmetry of the transformation under interchanging
coordinates and assuming that\ this function is kept by such a interchange,
it was shown that three different types of the transformation are possible.
The first type is a generalization of the Lorentz transformation. The second
and third types are principally different and possess unusual properties, in
particular, an uncertainty of time determination and solutions with lower
and upper frequency boundaries.

The point rotation frames have not transverse coordinates, however a
coordinate along the axis of rotation can be used as the space coordinate.
The approach developed in \cite{job} for two-dimensional (2D) case is
unacceptable for the 3D case. In this paper we somewhat modify the approach
using again only symmetry. The main idea of the modification is a "velocity
invariant" which remains unchanged by replacing any coordinate by another
and the use of the invariant for the construction of an additional
condition. The condition together with the condition of the speed of light
constancy defines the transformation parameters.

\section{Two-dimensional transformation}

In this section we demonstrate the approach on an example of the
two-dimensional transformation and show as the Lorentz transformation may be
deduced in such an approach.

The general form of the linear transformation for the transition from one
frame to another can be written as follows 
\begin{equation}
\tilde{z}=q(z-ut),\;\;\;\tilde{t}=\frac{\tilde{q}q-1}{\tilde{q}\tilde{u}}z-q%
\frac{u}{\tilde{u}}t,  \label{trp}
\end{equation}%
where $z$ and $t$ is an space coordinate and time, $u$ is the velocity of
the second frame relative to first one, the tilde corresponds to reverse
values. Eq. (\ref{trp}) turns out into the reverse transformation if
variables with the tilde change to variables without the tilde and vice
versa. Moreover the transformation is invariant by interchanging the
coordinates $z$ and $t$ correspondingly normalized$.$

In the general case $\tilde{u}$ and $u$ are connected by some dependence.
This dependence is assumed to be symmetric about $\tilde{u}$ and $u$: if $%
\tilde{u}$ is a function of $u$, then $u$ is the same function of $\tilde{u}$%
. Parameters $I_{1}=(\tilde{u}+u)/2,$ $I_{2}=(\tilde{u}-u)^{2}/4$ are
invariant under the reverse transformation therefore the dependence may be
rewritten in the form $I_{2}$ as some function of $I_{1}$ or vice versa.
Obviously the same is valid for another pair of parameters $I_{1}$ and $%
\tilde{u}u\equiv (I_{1}^{2}-I_{2}^{2})$ used early in \cite{job}. The
function is assumed to be differentiable infinite number of times and may be
presented as power series:

\begin{equation}
\text{ }I_{2}=\sum_{n=1}\lambda _{n}I_{1}^{n},\text{ \ or }%
I_{1}=\sum_{n=1}\lambda _{n}I_{2}^{n}  \label{ugen}
\end{equation}

In \cite{job} it was assumed that if to substitute into (\ref{ugen})
corresponding velocity obtained by interchanging coordinates then the
equality in (\ref{ugen}) is kept and this equality may be used as a
condition for the determination of $q$. However this approach cannot be used
for 3D case with the space and angle coordinate since interchange these
coordinates corresponds to replacing velocity by frequency. Below we modify
this approach. We form a "velocity invariant" keeping by such an interchange
and construct from this invariant an additional condition for $q$.

We start from the normalization of (\ref{trp}) as that the reverse velocity
is equal direct one with the opposite sign. An example of the normalization
is $z_{n}=z\sqrt{-\tilde{u}/u},$ \ $t_{n}=t,$ \ $\tilde{z}_{n}=\tilde{z}%
\sqrt{-u/\tilde{u}},$ \ $q_{n}=q\left( -\tilde{u}/u\right) ,$ $u_{n}=\sqrt{-%
\tilde{u}u}=-\tilde{u}_{n},$ where the subscript $"n"$ corresponds to a
normalized value.

Consider symmetry of the normalized transformation (\ref{trp}) about the
change of coordinates $z\rightarrow ht,t\rightarrow bz$, where $h,b$ are
dimensional coefficients. In the new variables the transformation has the
same form (\ref{trp}) if the new and old parameters are connected by
relations%
\begin{equation}
Q=q,\text{ }U=\frac{h}{b}\frac{\tilde{q}q-1}{\tilde{q}qu}\text{ },
\label{uq}
\end{equation}%
where $Q,U$ are new $q,u$ respectively.

The expression ("velocity invariant")%
\begin{equation}
I=bu+h\frac{\tilde{q}q-1}{\tilde{q}qu}  \label{I}
\end{equation}%
is kept under the above change. We may form two even invariants from the
velocity invariant: $I_{1}=\tilde{I}+I$, and $I_{2}=(\tilde{I}-I)^{2}$. The
invariants are kept under the sign change of $u$. In the given
two-dimensional case $\tilde{I}+I\equiv 0$ and, substituting $I_{1},$ $I_{2}$
into Eq. (\ref{ugen}) we find that the equation has a "fundamental solution" 
$I=0$. $h/b$ has the dimension of velocity squared. Let $h/b=c^{2}$,\ where $%
c$ is the speed of light, then the equality $I=0$ is non other than the
condition of the speed of light constancy. From the equality one follows%
\begin{equation}
\tilde{q}q=\frac{1}{1-u^{2}/c^{2}}.  \label{uq0}
\end{equation}%
We assume the equivalence of $\tilde{q}$ and $q$ that is $q$ must be an even
function of $u$: $\tilde{q}(u)\equiv q(-u)=q(u)$.

\section{Three-dimensional transformation}

The transformation in the 3-dimensional case may be written in the following
general form%
\begin{eqnarray}
\tilde{\varphi} &=&q_{1}[\varphi +p_{1}(z-ut)-\nu t],\text{ \ \ \ \ } 
\nonumber \\
\tilde{z} &=&q_{2}[p_{2}(\varphi -\nu t)+z-ut],\text{ \ \ \ \ \ }
\label{eq1} \\
\tilde{t} &=&q_{31}\varphi +q_{32}z+q_{33}t.  \nonumber
\end{eqnarray}%
We consider this transformation in application to electrooptics but assume
that such a transformation has more general character. In (\ref{eq1}) $%
\varphi $ is the angle between the electric vector of a plane circularly
polarized light wave and an direction in the first frame, $t$ is time, $\nu $
is the frequency of the second frame relative to first one, $u$ is the
velocity of the second frame relative to the first one, all parameters are
functions of $\nu ,u.$ The terms $p_{1}(z-ut)$ and $p_{2}(\varphi -\nu t)$
describe the velocity and frequency dismatch between the modulating and
modulated wave. It may be shown that parameters of the transformation are
connected by five independent equations: 
\begin{eqnarray}
q_{31} &=&\frac{1}{\tilde{u}^{\prime }}(\tilde{p}_{2}q_{1}+q_{2}p_{2}),\text{
\ \ }q_{32}=\frac{1}{\tilde{\nu}^{\prime }}(q_{1}p_{1}+\tilde{p}_{1}q_{2}),
\label{q23} \\
\text{\ }q_{33} &=&-\frac{\nu ^{\prime }q_{1}}{\tilde{\nu}}=\frac{u^{\prime
}q_{2}}{\tilde{u}},\text{ \ }(\tilde{q}_{1}-\frac{\nu }{u}\tilde{p}_{2}%
\tilde{q}_{2})(q_{1}-\frac{\tilde{\nu}}{\tilde{u}}p_{2}q_{2})=1,
\label{q1q1}
\end{eqnarray}%
where \ $\nu ^{\prime }=\nu +p_{1}u,$ \ $u^{\prime }=u+p_{2}\nu $. Except $%
\nu ,u$, the transformation (\ref{eq1}) has 7 parameters therefore 2 from
them are indeterminate and two extra conditions are necessary. The first
condition, as in the Lorentz case, is the speed of light constancy. The
second one follows from symmetry considerations.

Now we make normalization $\varphi _{n}=\zeta \varphi ,$ $z_{n}=\xi z,$ $\
q_{1n}=\tilde{\zeta}^{2}q_{1},$ $q_{2n}=\tilde{\xi}^{2}q_{2},$ $p_{1n}=$ $%
\zeta \tilde{\xi}p_{1},$ $p_{2n}=\tilde{\zeta}\xi p_{2},$ $\nu _{n}=\sqrt{-%
\tilde{\nu}\nu },$ $\tilde{\nu}_{n}=-\sqrt{-\tilde{\nu}\nu },\ u_{n}=\sqrt{-%
\tilde{u}u},$ $\tilde{u}_{n}=-\sqrt{-\tilde{u}u},$ 
\begin{equation}
\zeta =\sqrt{-\frac{\tilde{\nu}}{\nu }},\text{ \ \ }\tilde{\zeta}=\sqrt{-%
\frac{\nu }{\tilde{\nu}}},\text{ \ }\xi =\sqrt{-\frac{\tilde{u}}{u}},\text{
\ }\tilde{\xi}=\sqrt{-\frac{u}{\tilde{u}}}.  \label{nc}
\end{equation}%
Analogously to the Lorentz case we assume equivalence not only the direct
and reverse transformation but also the direct and opposite motion. It means
that in the normalized units $q_{k}(u,\nu )=$ $q_{k}(-u,-\nu )=q_{k}(-u,\nu
) $, where $k=1,2$.

\subsection{The speed of light constancy condition}

We use the speed of light constancy as one from conditions for the
definition of the transformation parameters. .

Consider circularly polarized light propagating through a medium with
rotating optical indicatrix (a single-sideband modulator \cite{sm},\cite{jpc}%
). Let $\omega =\varphi /t,$ $V=z/t$ to be the frequency and velocity of the
light wave in the first frame. If after passing the modulator the
polarization is kept then frequency and velocity at the output remains
unchanged. If the polarization is reversed $\tilde{\omega}\rightarrow -%
\tilde{\omega}$ (as it is in the single-sideband modulator at the half-wave
condition \cite{jpc}) then, using the direct and reverse transformation (\ref%
{eq1}), we can find the output velocity in the initial frame 
\begin{equation}
V^{\prime }=\frac{V-2\tilde{p}_{2}\tilde{q}_{2}q_{1}(\omega +p_{1}V-\nu
^{\prime })}{1-2\tilde{q}_{31}q_{1}(\omega +p_{1}V-\nu ^{\prime })},
\label{omp}
\end{equation}%
If $V$ equals the speed of light $c$ then $V^{\prime }=c$ and we obtain from
(\ref{omp}) the first condition%
\begin{equation}
q_{31}c=q_{2}p_{2}.  \label{c1}
\end{equation}%
Due to this condition the velocity in the second frame 
\begin{equation}
\tilde{V}=\frac{q_{2}(p_{2}\omega +V-u^{\prime })}{q_{31}\omega
+q_{32}V+q_{33}}  \label{omega}
\end{equation}%
equals $c$ as well. Moreover we can express all parameters in terms of $\nu
,u$ and $\gamma \equiv q_{2}/q_{1}$. For the determination of $\gamma $ an
additional condition is necessary.

\subsection{The second condition \qquad}

\ The transformation (\ref{eq1}) is kept under the change $(\varphi
,z,t)\rightarrow (a_{\varphi }\varphi ,h_{\varphi }t,b_{\varphi }z),$ $%
(\varphi ,z,t)\rightarrow (h_{z}t,b_{z}z,a_{z}\varphi ),$ $(\varphi
,z,t)\rightarrow (b_{t}z,a_{z}\varphi ,h_{z}t)$, where $a,b,h,$ with
corresponding indices are dimensional coefficients. Starting from $u$ and
using the change, we can construct the form ("velocity invariant")%
\begin{equation}
\rho \frac{q_{1}p_{1}}{q_{2}}+\tau \tilde{\nu}+\frac{1}{\rho }\frac{%
q_{2}p_{2}}{q_{1}}+\frac{1}{\beta }\tilde{u}+\frac{1}{\tau }\frac{q_{31}}{%
q_{1}}+\beta \frac{q_{32}}{q_{2}},  \label{form1}
\end{equation}%
where $\beta ^{2}\equiv h_{\varphi }/b_{\varphi },$ \ $\tau ^{2}\equiv
a_{z}/h_{z}\emph{,}$ \ $\rho ^{2}\equiv a_{t}/b_{t},$ $h_{t}^{2}=a_{t}b_{t},$
\ $b_{z}^{2}=a_{z}h_{z},$ \ $a_{\varphi }^{2}=b_{\varphi }h_{\varphi }$ and 
\begin{equation}
\rho =\beta \tau ,  \label{const}
\end{equation}%
The constants $\rho ,\tau ,\beta $ \ have the dimension of\ length, time and
velocity respectively. Analogously to the two-dimensional case we equate%
\[
\beta ^{2}=c^{2}. 
\]%
Exclude the constant $\rho $ with help of (\ref{const}), then in the
normalized units $\tau \nu \rightarrow \nu ,$ $u/\beta \rightarrow u$ the
invariant may be rewritten as%
\begin{equation}
I=\text{\ }\frac{\gamma ^{2}(1-u^{2})-1}{u^{2}\gamma \lbrack \gamma (1+u)+1]}%
[\nu -u(1+u)]+\frac{\nu (\gamma -1)}{u\gamma }-\nu -2u+\frac{2}{\nu }\frac{%
\gamma ^{2}(1-u^{2})-1}{[\gamma (1+u)+1]}.  \label{Inv}
\end{equation}%
where the positive value of $\beta $ is used. For negative $\beta $ the last
term in (\ref{Inv}) is excluded, this case is connected with negative $\rho $
or $\tau $, i.e., with the reflection of $z$ or $t$ at interchange
coordinates. In contrast to the two-dimensional case the invariant (\ref{Inv}%
) cannot be equal zero for arbitrary $\nu ,u$. Apparently the most probable
value of the characteristic constants $\tau $ is of the order of "nuclear"
time $\sim 10^{-23}\sec $. With this values normalized frequencies of the
electromagnetic field in the optical range are of the order of $10^{-8}\div
10^{-9}$.

Because of assumed $q_{k}(u,\nu )$ symmetry, $\gamma \equiv q_{2}/q_{1}$
also must keep under the reverse transformation $\tilde{\gamma}(u,\nu
)\equiv \gamma (-u,-\nu )=\gamma (u,\nu ).$ With this in mind we first
symmetrize the invariant (\ref{Inv}) with respect to the reverse
transformation%
\begin{equation}
I_{s}=\frac{1}{2}[\tilde{I}+I]=\frac{d^{\prime }}{u\gamma D}\nu -\frac{d}{%
\gamma D}-\frac{2}{\nu }\frac{d\gamma u}{D},
\end{equation}%
where 
\begin{equation}
d=\ \gamma ^{2}(1-u^{2})-1,\text{ }d^{\prime }=\ \gamma ^{2}(1+u^{2})-1,%
\text{ }D=d+2\gamma +2.  \label{dD}
\end{equation}%
Moreover we assume that $\gamma $ must be a even function under the sign
change of the velocity $\gamma (u,v)=\gamma (-u,v).$ The second condition
also must possess this property. Therefore we construct two even invariants. 
\begin{equation}
I_{1}\equiv \frac{1}{2}[I_{s}+I_{s}(-u)]=-\frac{d}{\gamma D},\text{ \ }%
I_{2}\equiv \frac{1}{4}[I_{s}-I_{s}(-u)]^{2}=\left( \frac{d^{\prime }\nu }{%
u\gamma D}-\frac{2}{\nu }\frac{d\gamma u}{D}\right) ^{2}  \label{Ie}
\end{equation}%
The invariants play the same role as in the two-dimensional case.
Accordingly to our approach we must postulate some dependence between $%
I_{1},I_{2}$ and consider this dependence as the second condition.
Analogously to the two-dimensional case we assume that this dependence is
differentiable infinite number of times and can be presented in the form of
a power series 
\begin{equation}
\left( \frac{d^{\prime }\nu }{u\gamma D}-\frac{2}{\nu }\frac{d\gamma u}{D}%
\right) ^{2}=\dsum\limits_{n=1}\lambda _{n}\left( \frac{d}{\gamma D}\right)
^{n}.  \label{IsIn}
\end{equation}%
The dependence $I_{1}$ equals power series in $I_{2}$ give the same results
and the choice of Eq. (\ref{IsIn}) is dictated by simplicity of \ the
expression.

Since coefficients $\lambda _{n}$ are unknown we restrict ourselves to the
expansion $\gamma $ in the vicinity of $u=0$ or $\nu =0$. Consider
parameters $p_{1},p_{2}$. Using the condition (\ref{c1}) we find from (\ref%
{q23}), (\ref{q1q1})%
\begin{equation}
\text{\ }p_{1}=\nu \frac{(\gamma ^{2}-1)(1+u)-\gamma u^{2}}{cu^{2}[\gamma
(1+u)+1]},\text{ \ }p_{2}=\frac{c}{\nu }\frac{\gamma ^{2}(1-u^{2})-1}{\gamma
\lbrack \gamma (1+u)+1]}.  \label{p1p2}
\end{equation}%
We require that \ $p_{1},p_{2}$ must be bounded at $u\rightarrow 0$.
Therefore the expansion of $\gamma $ in power series in $u^{2}$ is $\gamma
=1+\vartheta _{2}u^{2}+\vartheta _{4}u^{4}+\ldots ,$ where $\vartheta _{n}$
is a function of $\nu $. For $\vartheta _{2}$ we obtain from (\ref{IsIn})%
\begin{equation}
\vartheta _{2}=\frac{3-\sqrt{1-2\nu ^{2}/\lambda _{1}}}{2[1+\sqrt{1-2\nu
^{2}/\lambda _{1}}]}.  \label{teta2}
\end{equation}%
With this value the transformation (\ref{eq1}) in the limit $u\rightarrow 0$
has the form 
\begin{equation}
\tilde{\varphi}=\varphi +\nu \frac{1-\sqrt{1-2\nu ^{2}/\lambda _{1}}}{1+%
\sqrt{1-2\nu ^{2}/\lambda _{1}}}z-\nu t,\text{ \ \ }\tilde{z}=z,\text{ \ \ }%
\tilde{t}=t,\text{ \ \ \ \ \ \ }  \label{tru}
\end{equation}%
where the normalized values $z\rightarrow z/\rho ,$ $t\rightarrow t/\tau $
are used. For the positive $\lambda _{1}$ values of $\nu ^{2}$ have the
upper boundary $\nu ^{2}\leq \lambda _{1}/4.$

From the other hand if $\nu =0$ then the solution of Eq. (\ref{IsIn}) $d=0$
with $\gamma =1/\sqrt{1-u^{2}}$ exist. This value may be used as the first
term of the expansion $\gamma =1/\sqrt{1-u^{2}}+\gamma _{2}\nu ^{2}+\gamma
_{4}\nu ^{4}+\ldots ,$ where $\gamma _{n\text{ \ }}$is a function of $u.$ In
this case%
\begin{equation}
\gamma _{2}=\frac{1}{2\sqrt{1-u^{2}}}\frac{-1+\sqrt{1+8u^{2}/\lambda }}{1+%
\sqrt{1+8u^{2}/\lambda }},  \label{gam2}
\end{equation}%
where $\lambda =\lambda _{1}(1-u^{2})(\sqrt{1-u^{2}}+1)$. If $u\rightarrow 0$
and $u\rightarrow 1,$ then $\gamma _{2}\rightarrow u^{2}/2\lambda _{1},$
and\ \ $\gamma _{2}\rightarrow 1/(2\lambda _{1}\sqrt{1-u^{2}})$. With the
solution \{\ref{gam2}) the transformation (\ref{eq1}) exactly corresponds to
the Lorentz transformation in the limit $\nu \rightarrow 0$%
\begin{equation}
\tilde{\varphi}=\varphi ,\text{ \ \ }\tilde{z}=\frac{z-ut}{\sqrt{1-u^{2}}},%
\text{\ \ \ }\tilde{t}=\frac{-uz+t}{\sqrt{1-u^{2}}}.  \label{trnu}
\end{equation}%
The solution in the form of an expansion of $\gamma $ in a power series in $%
\nu ^{2}$ represents a generalization of the Lorentz transformation. It may
be straightforwardly shown that if the characteristic time $\tau $ is of the
order of "nuclear time" $\sim 10^{-23}sec$ then the generalized Lorentz
transformation as well as the transformation in the form (\ref{tru}) cannot
give an extra frequency shift by the single-sideband modulation discussed in 
\cite{job}.

Numerical calculations of $\gamma $ for different dependences $%
I_{2}=I_{2}(I_{1})$ as in the form of powers series as other forms\
demonstrate a whimsical variety of solutions.

However the question on the explicit form of this dependence remains open.

\section{Conclusion}

We have considered 3D transformation for the point rotation coordinate
frames with coordinates angle, length and time. Main problem of such a
transformation is two additional conditions for the definition of the
transformation parameters and the expression of parameters as functions of
velocity and frequency. The speed of light constancy can be used as one
condition. However there is no a principle like relativity principle for the
second condition. Instead we use a "velocity invariant". The invariant,
constructed from parameters corresponding to velocity at every interchanging
the coordinates, is kept by such an interchange as well as the
transformation itself. In approach considered in the paper two even
invariants are extracted from this invariant. A dependence between these
invariants defines the second condition. This dependence is not completely
definite therefore the transformation may be investigated only in vicinity
of zeroth velocity or frequency. If the frequency tends to zero then in this
limit the transformation coincides with the Lorentz transformation.\ \ \ \ \
\ \ \ \ \ \ \ \ \ \ \ \ \ \ \ \ \ \ \ \ \

\end{document}